\documentclass{aa}
\usepackage[varg]{txfonts}
\usepackage{xcolor}
\usepackage{graphicx}
\usepackage{siunitx}
\usepackage{dcolumn}
\usepackage{ulem}
\usepackage[export]{adjustbox}
\usepackage{multirow}
\usepackage[skip=1ex]{caption}
\usepackage{subcaption}

\newcolumntype{d}{D{.}{.}{-1}}

\newcommand{\pa}{\ensuremath{_{\parallel}}}
\newcommand{\se}{\ensuremath{_{\perp}}}
\newcommand{\md}{\ensuremath{\mathrm{d}}}
\providecommand\gt{\ensuremath{>}}
\providecommand\lt{\ensuremath{<}}

\allowdisplaybreaks
\graphicspath{{Figures/}}

\begin{document}

\title{Revisiting the Ulysses electron data with a triple fit of velocity distributions}

\author{K. Scherer\inst{\ref{inst1},\ref{inst2}}
  \and E. Husidic\inst{\ref{inst3},\ref{inst4}} \and M. Lazar\inst{\ref{inst3}}  \and H. Fichtner\inst{\ref{inst1},\ref{inst2}}}

\institute{Institut f\"ur Theoretische Physik, Lehrstuhl IV:
  Plasma-Astroteilchenphysik, Ruhr-Universit\"at Bochum, D-44780 Bochum,
  Germany \email{kls@tp4.rub.de}\label{inst1}
  \and
Research Department, Plasmas with Complex Interactions,
  Ruhr-Universit\"at Bochum, 44780 Bochum, Germany\label{inst2}
  \and
  Centre for mathematical Plasma Astrophysics, Department of Mathematics, KU Leuven, Celestijnenlaan 200B, 3001 Leuven, Belgium\label{inst3}
  \and
  Department of Physics and Astronomy, University of Turku, 20014 Turku, Finland\label{inst4}
}

\abstract
{Given their uniqueness, the Ulysses data can still provide us with valuable new clues about the properties of plasma populations in the solar wind, and, especially, about their variations with heliographic coordinates. In the context of kinetic waves and instabilities in the solar wind plasma, the electron temperature anisotropy plays a crucial role. To date, mainly two electron populations (i.e., the core and the halo) have been surveyed using anisotropic fitting models, and limiting in general to the ecliptic observations.}
{We revisit the electron data reported by by the SWOOPS instrument on-board of the Ulysses spacecraft between 1990 to early 2008. 
These observations reveal velocity distributions out of thermal equilibrium, with anisotropies (e.g., parallel drifts or/and different temperatures, parallel and 
perpendicular to the background magnetic field), and quasi-thermal and suprathermal populations with different properties.}
{We apply a 2D nonlinear least square fitting procedure, using the Levenberg-Marquardt algorithm, to simultaneously fit the velocity electron data (up to a few keV) with a triple model 
combining three distinct populations: the more central quasi-thermal core and suprathermal halo, and a second
suprathermal population consisting mainly of the electron strahl (or beaming population with a major field-aligned drift).
The recently introduced $\kappa$-cookbook is used to describe each component with the following anisotropic distribution functions (recipes): Maxwellian, regularized $\kappa$-, and generalized $\kappa$-distributions.
Most relevant are triple combinations selected as best fits (BFs) with minimum relative errors and standard deviations. }
{The number of BFs obtained for each fitting combination sum up to 80.6 \% (70.7 \% in the absence of coronal mass ejections) of the total number of events. 
Showing the distribution of the BFs for the entire data set, during the whole interval of time, enables us to identify the most representative fitting combinations associated with either fast or slow winds, and different phases of solar activity.
The temperature anisotropy quantified by the best fits is considered as a case study of the main parameters characterizing electron populations. 
By comparison to the core, both suprathermal populations exhibit higher temperature anisotropies, which slightly increase with the energy of electrons. 
Moreover, these anisotropies manifest different dependencies on the solar wind speed and heliographic coordinates, and are highly conditioned by the fitting model.}
{These results demonstrate that the characterization of plasma particles is highly dependent on the fitting models and their combinations, and this method must be considered with caution. The multi-distribution function fitting of velocity distributions has however a significant potential to advance  understanding of the solar wind kinetics and deserves further quantitative analyses.}

\maketitle

\section{Introduction}\label{sec:introduction}

The solar wind is a hot and dilute plasma that constantly streams from the Sun and 
fills interplanetary space \citep{Marsch-2006, Lazar-2012}.
Its collisionpoor nature allows for  departures from thermal (Maxwellian) equilibrium 
\citep{Kasper-etal-2006, Stverak-etal-2008, Wilson-etal-2019b, Wilson-etal-2020}, which persist, being most probably maintained by the the resonant interaction with wave turbulence and fluctuations
\citep{Bale-etal-2009, Yoon-2011, Alexandrova-etal-2013}. 
In-situ observations regularly reveal typical non-thermal characteristics in the particles' velocity
distributions including the following: (i) enhanced supra-thermal tails caused by an increased number of particles in 
the high-energy regime of the distribution \citep{Maksimovic-etal-1997, Stverak-etal-2008,Mason-Gleckler-2012}, 
(ii) temperature anisotropies, that is, different temperatures parallel and perpendicular
to the ambient magnetic field \citep{Kasper-etal-2006, Marsch-2006,Stverak-etal-2008}, 
and (iii) anti-sunward field-aligned beams (also called strahls) 
\citep{Pilipp-etal-1987a, Pierrard-etal-2001, Wilson-etal-2019a}.

In the electron distributions up to a few keV, three prominent components can be identified
\citep{Pierrard-etal-2001, Wilson-etal-2019a}. First, the core of the distribution is represented by 
a quasi-thermal component, with up to 80 to 90 \% of the total particle number density, and well described by a Maxwellian distribution. Second, with about 5 to 10\% of the total number density, 
a suprathermal component, commonly referred to as halo, contributes to an enhancement of the distribution tails and can 
be modeled by an Olbertian Kappa (or $\kappa$--) power-law distribution function. Third, a further constituent termed beam or strahl can be found and has a noticeable field-aligned drift (or relative beaming
speed) \citep{Pilipp-etal-1987a, Marsch-2006}. The number density of the strahl population is even less than that of the halo, and can also be described by a (drifting) $\kappa$-distribution \citep{Wilson-etal-2019a}. 
While these three components can individually be described by 
the mentioned distribution functions, different combinations
are employed for an overall fit. If the beam has a very low density, a dual model can be used to fit the core and halo \citep{Lazar-etal-2017}. Otherwise, if the beam 
appears more clearly, a triple model that includes the strahl can be invoked \citep{Wilson-etal-2019a}. In closed magnetic field topologies, such as coronal loops, where double strahls, that is, two counterbeaming strahls have been observed, see \cite{Lazar-etal-2014} and references therein, a quadruple model can be applied \citep{Macneil-etal-2020}. 
Sometimes a superhalo component is mentioned  \citep{Yoon-etal-2013}, but these populations may enhance the higher energy tails above 10~keV \citep{Lin-1998}.  
Thus, with the electron data up to a few keV, and excluding those associated with closed magnetic field lines of coronal mass ejections, the present study is limited to a triple model (see Sec.~\ref{sec:models}).
 
Particles in heliospheric plasmas such as the solar wind are subject to processes involving non-thermal
acceleration. Their distribution tails then no longer exhibit a Maxwellian (i.e., exponential) cutoff, 
but often a decreasing power law. Such non-equilibrium distributions are well parameterized
by the Kappa distribution, introduced empirically by \cite{Olbert-1968}, and published 
for the first time by \cite{Vasyliunas-1968} as a global fitting model that does not distinguish between
core and halo. More rigorous analyses involve a combination of multiple (anisotropic) distribution functions, 
including Maxwellian and Kappa distributions
 \citep{Pilipp-etal-1987b,Pilipp-etal-1987c,Pilipp-etal-1987a,Maksimovic-etal-2005, Stverak-etal-2008}.
The Kappa distribution did not only prove to be a powerful tool for modeling non-thermal distributions,
but has also become notorious for its critical limitation in defining macroscopic physical properties by the velocity moments, for example of order $l$, which diverge for low power exponents $\kappa < (l +1)/2$   \citep{Lazar-Fichtner-2021}. For this reason, a generalization of the (isotropic) standard Kappa distribution has recently been
introduced by \cite{Scherer-etal-2017}, termed the regularized Kappa distribution, for which all 
velocity moments converge. An extension to the anisotropic regularized Kappa distribution was then presented in \cite{Scherer-etal-2019b}. The mathematical definitions of these distribution functions are given 
in Sec.~\ref{sec:models}.

The present paper aims at a re-evaluation of the Ulysses electron data obtained between 1990 and 2008, and is building on the work in \cite{Scherer-etal-2021}.
For a realistic analysis, we incorporated the anisotropic nature of the distributions by applying a 2D fitting method. 
In order to take potential single components of the distributions into account, we use a triple model including a quasi-thermal core, a suprathermal halo, and a suprathermal strahl component. 
For the model distributions, we chose the  anisotropic bi-Maxwellian, the regularized bi-Kappa, and the generalized anisotropic regularized Kappa distribution, which was introduced in an attempt to unify the various commonly used Kappa distributions \citep{Scherer-etal-2021}.
By establishing conditions defining good fits and best fits, in  section~\ref{sec:models} we describe their distributions for the entire data set, and for each year in part.  Section~\ref{sec:time_speed_var} contains a breakdown of the Ulysses data according
to latitude and solar wind speed, and establish a connection to the point in the solar cycle at that time.
After introducing formulas used to compute electron parameter (e.g., temperature anisotropies), in section~\ref{sec:parameters} we consider temperature anisotropy as a case-study, and identify correlations between temperature anisotropy and other quantities such as the solar wind speed, parallel plasma beta and other parameters more specific to distribution models. 
The paper ends with conclusions in Sec.~\ref{sec:conclusions}. 
%
\section{The models} \label{sec:models}

We fit the Ulysses electron data from the SWOOPS instrument
\citep[][\footnote{ \url{http://ufa.esac.esa.int/ufa/\#data}}]{Bame-etal-1992} (in "Additional datasets")
by assuming that there exist up to three electron populations: a core
component indicated by the subscript $c$,
a halo component by the subscript $h$,
and a strahl component by the subscript $s$.
The total combined distribution function and its moments are indicated
by the subscript $t$:
\begin{align}
  f_{t} = f_{c} + f_{h} + f_{s}\,,
\end{align}
where the distribution functions $f_{i}$, $i  \in \{c,h,s\}$ are described below. We
seek to satisfy the following condition:
\begin{align}\label{eq:ncond}
  n_{c} > n_{h} > n_{s}
\end{align}
with $n_{i}$, $i \in \{c,h,s\}$ denoting corresponding number
densities.

The distribution function $f_{i}$ with $i \in \{c,h,s\}$ can be an
anisotropic Maxwellian distribution (AMD), a regularized anisotropic
Kappa-distribution (RAK) or a generalized anisotropic
Kappa-distribution (GAK). These types of distributions can be described
by the recipes introduced in \citet{Scherer-etal-2020c}, where the
general recipe $(\eta_{\parallel},\eta_{\perp},\zeta,
\xi_{\parallel}, \xi_{\perp})$, abbreviated already with GAK, is given by
\begin{align}\label{eq:GAK}
   &f_{GAK}(\eta\pa,\eta\se,\zeta,\xi\pa,\xi\se)
  = n_{0}\,N_{{GAK}}\\\nonumber
      &\times \left(1+ \frac{(v_\parallel - u)^2}{\eta\pa\Theta\pa^{2}} +
      \frac{v\se^{2}}{\eta\se\Theta\se^{2}}\right)^{-\zeta}
      \exp\left({-\xi\pa\frac{(v_\parallel - u)^2}{\Theta\pa^{2}}-\xi\se\frac{v\se^{2}}{\Theta\se^{2}}}\right)\,,
\end{align}
where $v_\parallel$ and $v_\perp$ are the parallel and perpendicular velocity components, respectively, with respect
to the magnetic field, and $u$ is a parallel drift speed. The parameters $\eta\pa,\eta\se,\zeta,\xi\pa,\xi\se$ are constants with respect
to velocity, space and time, and 
$\Theta\pa$ and $\Theta\se$ normalize the velocity components and often are termed thermal speeds.
The normalization constant of the distribution function in Eq.~\eqref{eq:GAK} reads
\begin{align}\label{eq:nag}
  N_\mathrm{GAK}^{-1} = &\sqrt{\pi^{3}}
                          \Theta\pa\Theta\se^{2}\eta\pa^{\frac{1}{2}}\eta\se \\\nonumber
                        &\times \int\limits_{0}^{1}U\left(\frac{3}{2},\frac{5}{2}-\zeta,
                          \xi\se\eta\se +  (\xi\pa\eta\pa-\xi\se\eta\se)t^{2}\right) \md t
\end{align}
with $U(a,b,x)$ being the Kummer-U function. 
The AMD is then given by the recipe $(1,1,0,1,1)$, while the RAK is represented by
$(\kappa,\kappa,\kappa+1,\xi_{\parallel},\xi_{\perp})$ To these dependent variables come in addition the dependent variables $n_0,\Theta_{\pa},\Theta_{\se}$, and $u$. Thus, for the AMD we have to fit 4 parameters, for the GAK 9 and for the RAK 7. We allowed that the number of data points equals the number of free parameters, which is rarely the case. Usually there are more than 60 data points to be fitted. If there where less, the mean error and the standard deviation are larger than 0.3 (see below).

\begin{figure}[t!]
\centering
\includegraphics[width=1.0\columnwidth]{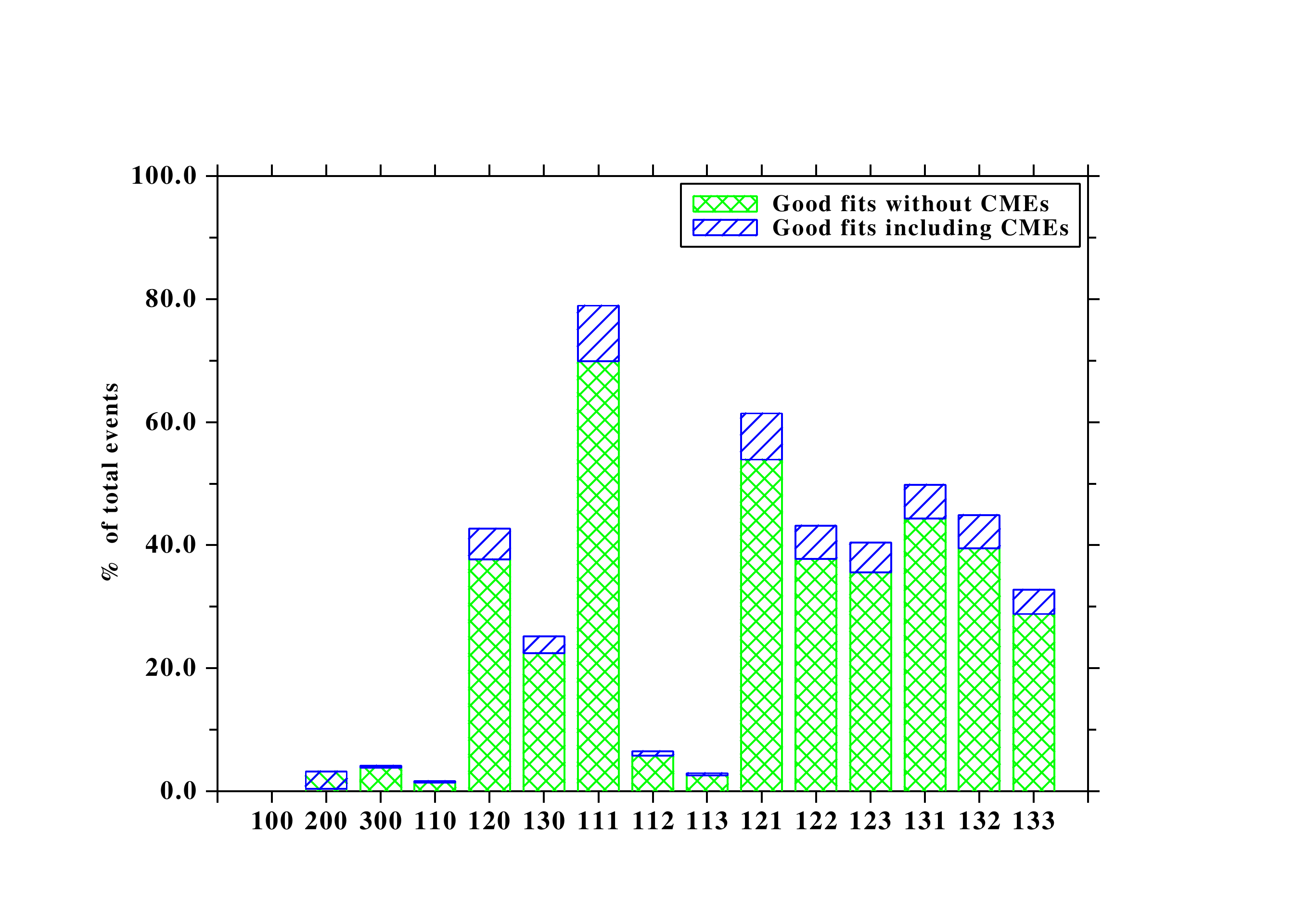}
\caption{Histogram of the GFs of the entire data set for each fitted (total) distribution $f_{ijk}$. Blue boxes show the additional GFs for the events during CMEs. See Table~\ref{t1} for the corresponding values.} \label{fig:1}
\end{figure}

For the total combined distribution function we introduce the generic notation $f_t = f_{ijk}$, where this time indices $i, j$ and $k$ indicate the fitting models. Thus, to avoid further clumsy notation, we use the index $1$ for the AMD,
$2$ for the GAK, and $3$ for the RAK, while $0$ indicates that no model is used for the respective component (i.e., for singular or dual fitting models). We write, for
example, for a distribution function
\begin{align}
  f_{123} = f_{c,1} + f_{h,2} + f_{s,3}\,,
\end{align}
which means that $f_{123}$ has an AMD core, a GAK halo and an RAK
superhalo/strahl.
We used the fitting method described in
\citet{Scherer-etal-2021}, which is similar to the one applied by
\citet{Wilson-etal-2020}. The data were fitted with the following combinations of distribution
functions: $f_{100}$, $f_{200}$, $f_{300}$, $f_{110}$, $f_{120}$,  $f_{130}$,
$f_{111}$, $f_{112}$, $f_{113}$, $f_{121}$, $f_{122}$, $f_{123}$, $f_{131}$,
$f_{132}$, $f_{133}$. For dual or triple combinations, we always assume that the core distribution is Maxwellian.

\begin{figure}[t!]
  \centering
  \includegraphics[width=1.0\columnwidth]{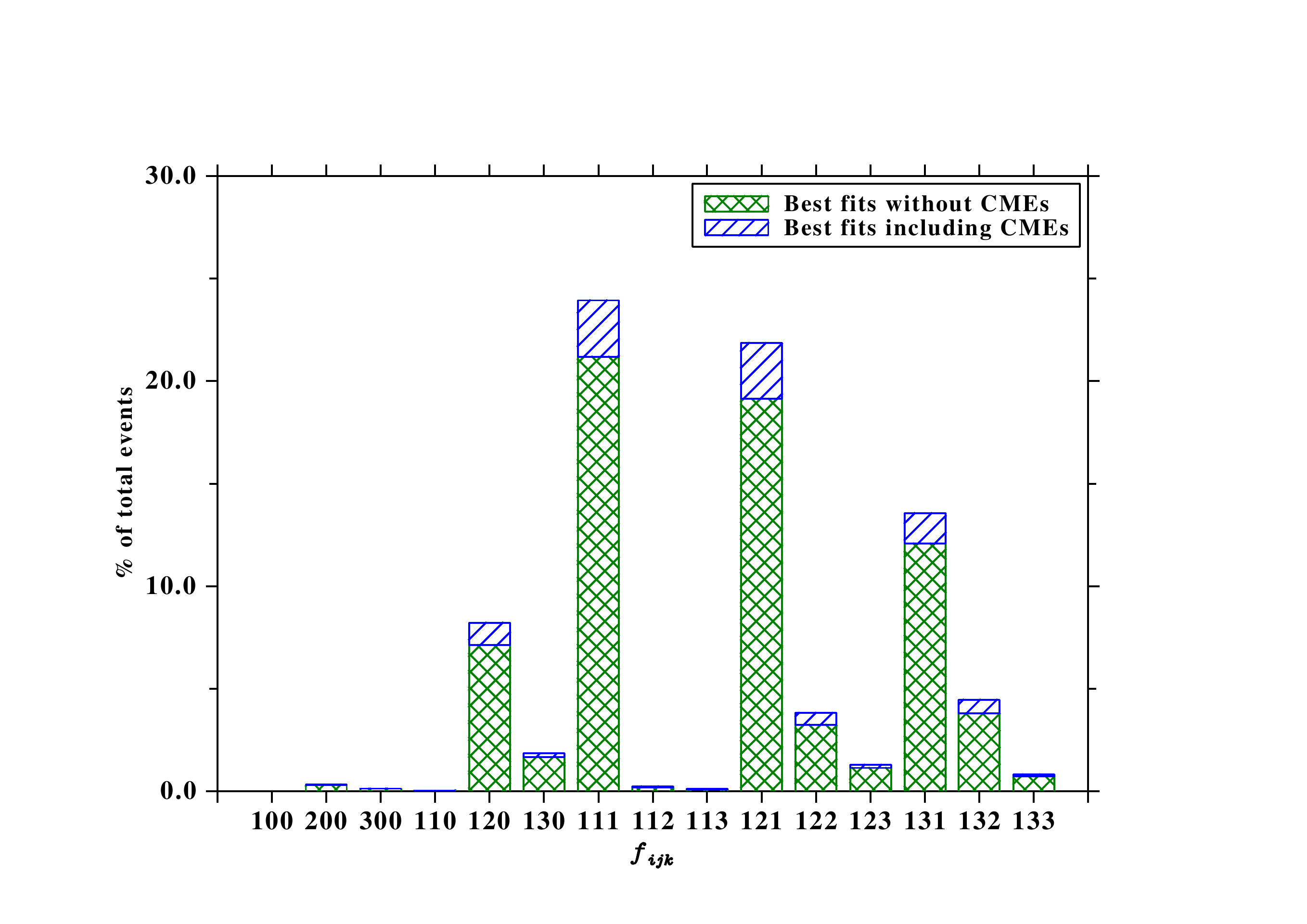}
\caption{Histogram for the BFs of the entire data set. The green boxes show the BFs without
CMEs, the blue boxes show the additional BFs with CMEs. See Table~\ref{t1} for the corresponding values.}\label{fig:2} 
\end{figure}

To check the quality of the fits we  define the relative error
 \begin{align}
E_{i} =\frac{|f_{fit,i}-f_{obs,i}|}{f_{obs,i}}
 \end{align}
 for each data point $i$,  and the mean relative error $<E>$ and  its standard deviation $\sigma$  as
 \begin{align} 
<E> & = \frac{1}{N} \sum\limits_{i=1}^{N} E_i \,, \label{eq:mean} \\ 
\sigma & =\sqrt{\frac{1}{N-1} \sum\limits_{i=1}^{N} (<E> - E_i)^2} \,. \label{eq:sigma}
\end{align}
For a good fit (GF), see Fig.~\ref{fig:1}, we require that $<E> \; \le 0.3$ and $\sigma \le 0.3$
\citep{Scherer-etal-2021}. 
The events which do not obey this condition or condition~\eqref{eq:ncond} for the
number densities are rejected.
Furthermore, we define the best fit (BF), see Fig.~\ref{fig:2}, as
the minimum of $<E>$ and $\sigma$ of all GFs. 

Our analysis covers the Ulysses data from the launch in late 1990 to early  2008.
There are in total 324,450 events, including 30,558 events during coronal
mass ejections (CMEs), which are taken from \citet{Richardson-2014}. During CMEs (reduced
in number, below 10~$\%$ of the number of events of a certain relevance) the electrons
may exhibit a double strahl or two beaming components, sunward and anti-sunward, moving
along the closed magnetic field topology. Thus, a double strahl (more or less symmetric) is not reproduced by our models, but it can
mimic a suprathermal component with an excess of temperature anisotropy in the direction parallel to the magnetic field. To avoid
such a confusing interpretation, the events during CMEs are counted separately in our present analysis. 
In addition, we have rejected 2,139 events, which violated condition~\eqref{eq:ncond}.  The total
number of bad fits is about 20\%. Ideally, a data set consists of 400 data points with
finite values, but most of the time there are much fewer such points, due to the missing data.
If the number of these points is too low, the fits become unreliable, as indicated
by the mean error and the standard deviation in the case of rejected fits.  The data set is given in keV without any error estimates, thus we were not able to weight the data according to their observational errors, and therefore the weight is always unity.

In Fig.~\ref{fig:1} we show the distribution of GFs for all individual distributions functions $f_{ijk}$.
The most reduced relevance can be attributed to singular fits, like $f_{100}$,
$f_{200}$ or $f_{300}$, with less than $4~\%$ of total events, and those reproducing the halo with a Maxwellian and the strahl with GAK
($f_{112}$ with GFs $<7~\%$) or RAK ($f_{113}$, with GFs $<3~\%$); see Table~\ref{t1}. The
highest peak is given by a standard combination of three AMDs, that is, the $f_{111}$ combination, with the ability to provide GFs for about 70 $\%$ of
the total data (in the absence of CMEs, with green color). However, GFs are also obtained with all the other combinations involving GAK
or RAK for describing the suprathermal populations. These combinations dominate the histogram with representations between 20~\% and
more than 50~$\%$ of the total number of events: $f_{130}$ with $\sim 22.5~\%$, $f_{133}$ with $\sim 29~\%$, $f_{120}$,
$f_{122}$ $f_{123}$ and $f_{132}$, each of them approaching $40~\%$, $f_{131}$ with $\sim 45~\%$, and $f_{121}$ with $\sim 54~\%$; see
again Table~\ref{t1}. 

\begin{table}
\caption{Number of events (in $\%$) corresponding to the BFs and to the GFs of all combinations of distribution functions in Figs.~\ref{fig:1} and \ref{fig:2}.} \label{t1}
\centering
  \begin{tabular}{l|dddd}
    $f_{ijk}$  & \multicolumn{1}{c}{BF  w/o CMEs} & \multicolumn{1}{c}{BF} & \multicolumn{1}{c}{GF w/o CMEs} & \multicolumn{1}{c}{GF}\\
    \hline
$f_{100}$ &  0.0 &  0.0 &  0.0  & 0.0 \\
$f_{200}$ &  0.3 &  0.3 &  3.2  & 0.3 \\
$f_{300}$ &  0.1 &  0.1 &  3.9  & 4.2 \\
$f_{110}$ &  0.0 &  0.0 &  1.5  & 1.7 \\
$f_{120}$ &  7.1 &  8.2 & 37.7  &42.7 \\
$f_{130}$ &  1.7 &  1.9 & 22.4  &25.2 \\
$f_{111}$ & 21.2 & 23.9 & 69.9  &78.9 \\
$f_{112}$ &  0.2 &  0.2 &  5.8  & 6.5 \\
$f_{113}$ &  0.1 &  0.1 &  2.6  & 2.9 \\
$f_{121}$ & 19.1 & 21.9 & 53.9  &61.4 \\
$f_{122}$ &  3.2 &  3.8 & 37.7  &43.2 \\
$f_{123}$ &  1.1 &  1.3 & 35.6  &40.4 \\
$f_{131}$ & 12.1 & 13.6 & 44.3  &49.8 \\
$f_{132}$ &  3.8 &  4.5 & 39.5  &44.9 \\
$f_{133}$ &  0.7 &  0.8 & 28.8  &32.7 
\end{tabular}
\end{table}

These results are refined in Fig.~\ref{fig:2}, which shows the distribution of the BFs for the entire data set (with green for the events
without CMEs). The number of BFs (in $\%$) obtained for each combination of distribution functions are given in Table~\ref{t1}, and together sum up to
80.6~\% (70.7~\% without CMEs) of the total number of events. Thus, each valid event is assigned one of these fitting combinations $f_{ijk}$.
It can be seen that 21.2~\% of the events  without CMEs are best fitted by the $f_{111}$ combination. This is followed at short difference by the
$f_{121}$ with 19.1~\%, and then by $f_{131}$ with 12.1\%, $f_{120}$ with 7.1~\%, $f_{132}$ with 3.8~\%, $f_{122}$ with 3.2~\%, and $f_{130}$ with
1.7~\%. Remarkable is the existence of dual core-halo distributions (in the absence of strahl), i.e., $f_{120}$ and $f_{130}$, with almost 9~\% of the
relevant events. BFs around 1~\% are obtained for combinations like $f_{123}$ and $f_{133}$, while singular distributions using GAK (i.e., $f_{200}$)
or RAK (i.e., $f_{300}$) have a very reduced presence, with 0.3~\% and 0.1~\%, respectively. 
Fitting models involving generalized Kappa distributions, like GAK or RAK, sum up together to 56.7~\% of the events (49.5~\% without CMEs). 
It is noteworthy that 59.4~\% of the events (52.4~\% without CMEs) have the strahl component best reproduced by an AMD, while halos are described by a GAK in about 35~\% of the events (30.5~\%
without CMEs), and by an RKD in about 21~\% of the events (18.3~\% without CMEs). We must also outline that the most prominent fit combines three AMDs, for 24\% of all events (21.2~\% without CMEs). 

\begin{figure} [t!]
  \centering
   \includegraphics[width=1.0\columnwidth]{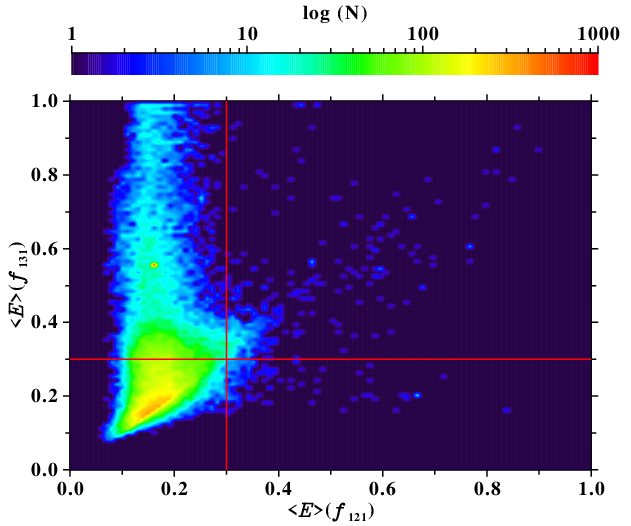}
   \caption{Correlating the mean error of GFs of the  $f_{121}$ distribution and the mean error of GFs of the $f_{131}$ distribution. The red line indicates the threshold for a GF.} \label{fig:20}
 \end{figure}

In order to check if the more complicated GAK-halo distribution ($f_{121}$) could be replaced by a simpler RAK-halo distribution 
($f_{131}$), we compare the BFs of the $f_{121}$ distribution with the GFs of the $f_{131}$ distribution.
Figure~\ref{fig:20} displays the mean error of the BFs of the $f_{121}$ distribution against the mean error of GFs of the
corresponding $f_{131}$ distribution. It can be seen that about half of the $f_{121}$ distributions can be replaced by the
$f_{131}$ distribution and obtaining still a GF. In total we have 63,821 $f_{121}$ BF events, which can be replaced by 37,263 GF
events of the $f_{131}$ distribution. Thus, about 58\% of the $f_{121}$ distribution could be replaced by the $f_{131}$
distribution. This can be helpful in analytic studies of dispersion relations. Nevertheless, we still have about 42\%
(26,558) events of the $f_{121}$ distribution for which the fits of the $f_{131}$
distribution give only bad results.

The total number of bad fits is about 19.4\% or 62,843 events, together with 2,139 rejected events (0.66\%). These data shall be handled individually
(about 20\%), because the fit procedure needs to start with a different initial guess, the data sets are too spare to be fitted, or the data cannot be
fitted with the above combinations of distribution functions. In the present analysis, from the total number of events we consider the remaining majority of 80~\%, or 259,365 events.
In the following we discuss the correlation between macroscopic parameters, and concentrate only on the BFs, which are unique. We mainly refer to the most representative combinations like $f_{111}, f_{121}$, and $f_{131}$, although the analysis may also take into account the less prominent examples like $f_{120}, f_{122}$, or $f_{132}$.

\begin{figure*} [h!]
 \centering
  \includegraphics[width=1.8\columnwidth]{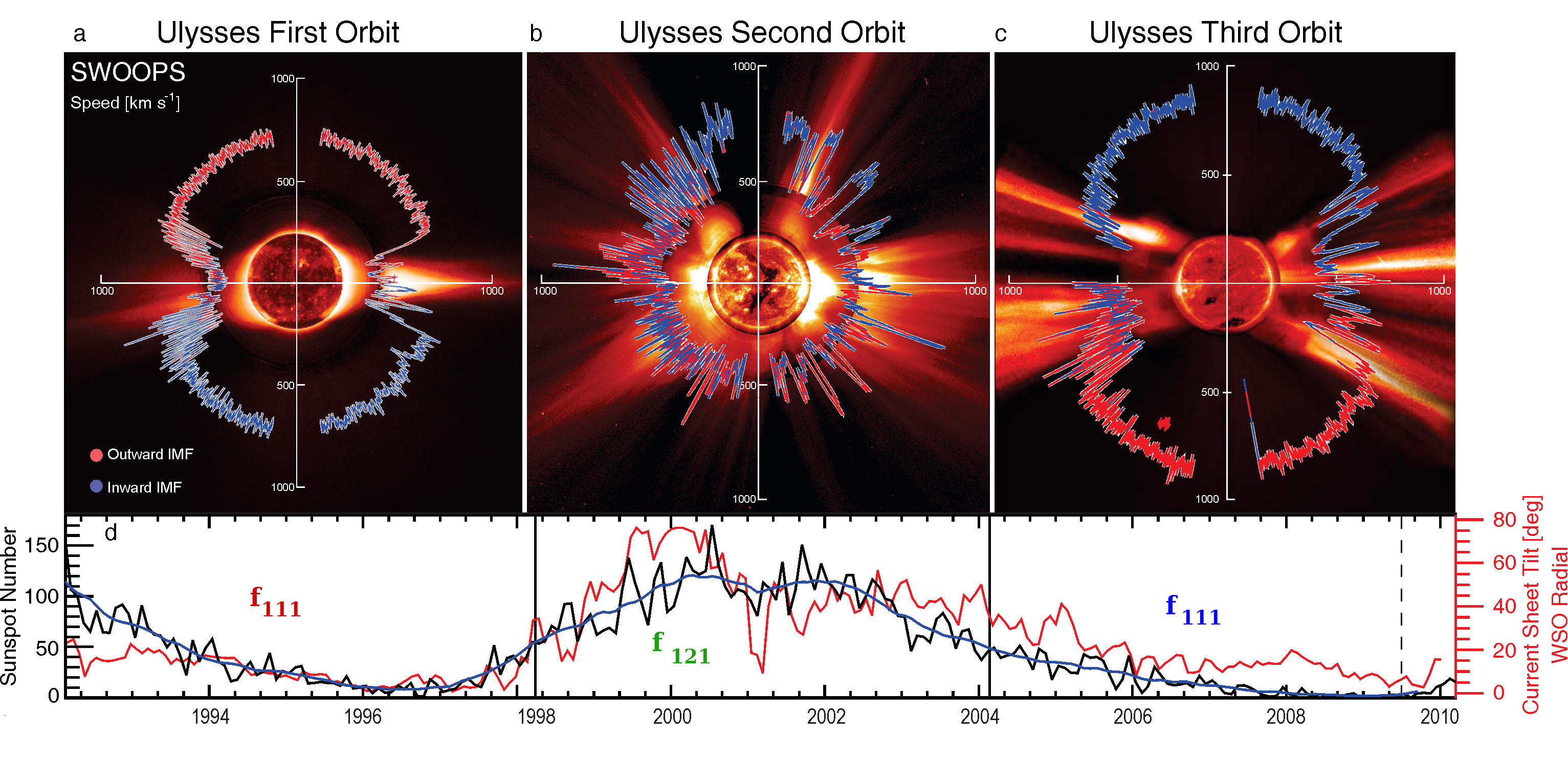}
 \caption{The Ulysses solar wind speeds during time. Adapted from \citet{McComas-2008}.  The three upper panels show the solar wind proton speed as a function of heliographic latitude in a polar coordinate system, during solar minimum (right upper panel), solar maximum (middle upper panel) and again solar minimum (left upper panels).}  The corresponding sun spot number (solar activity) is shown in the lower panel, where the black line gives the sunspot number and the red line the tilt angle. Indicated in the lower inlet are the dominant distribution functions during that period to illustrate the solar cycle dependence of the electron distribution functions as shown in Fig.~\ref{fig:3}. 
  \label{fig:0}
\end{figure*}

\begin{figure} [h!]
 \centering
  \includegraphics[width=0.9\columnwidth]{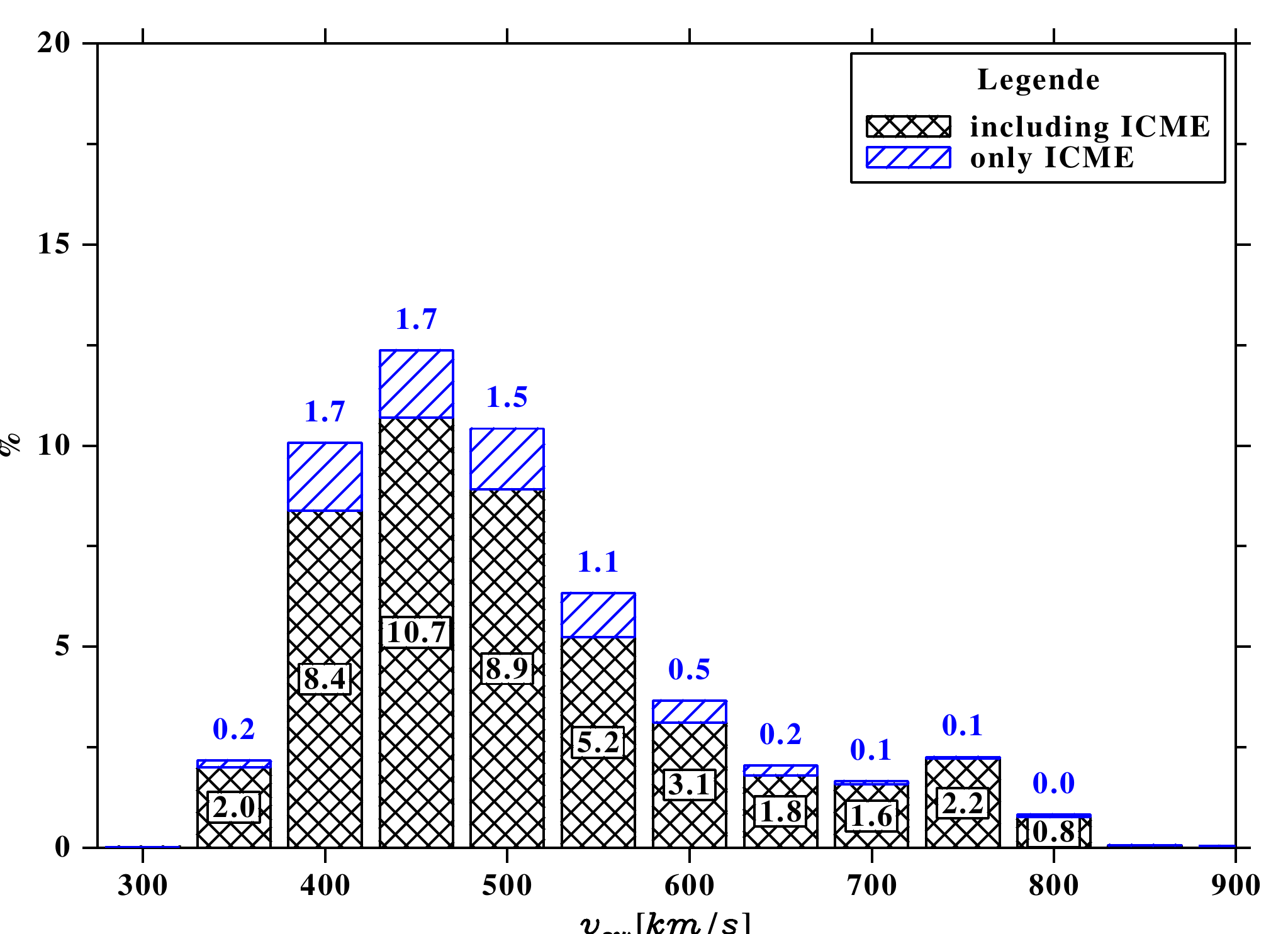}
  \includegraphics[width=0.9\columnwidth]{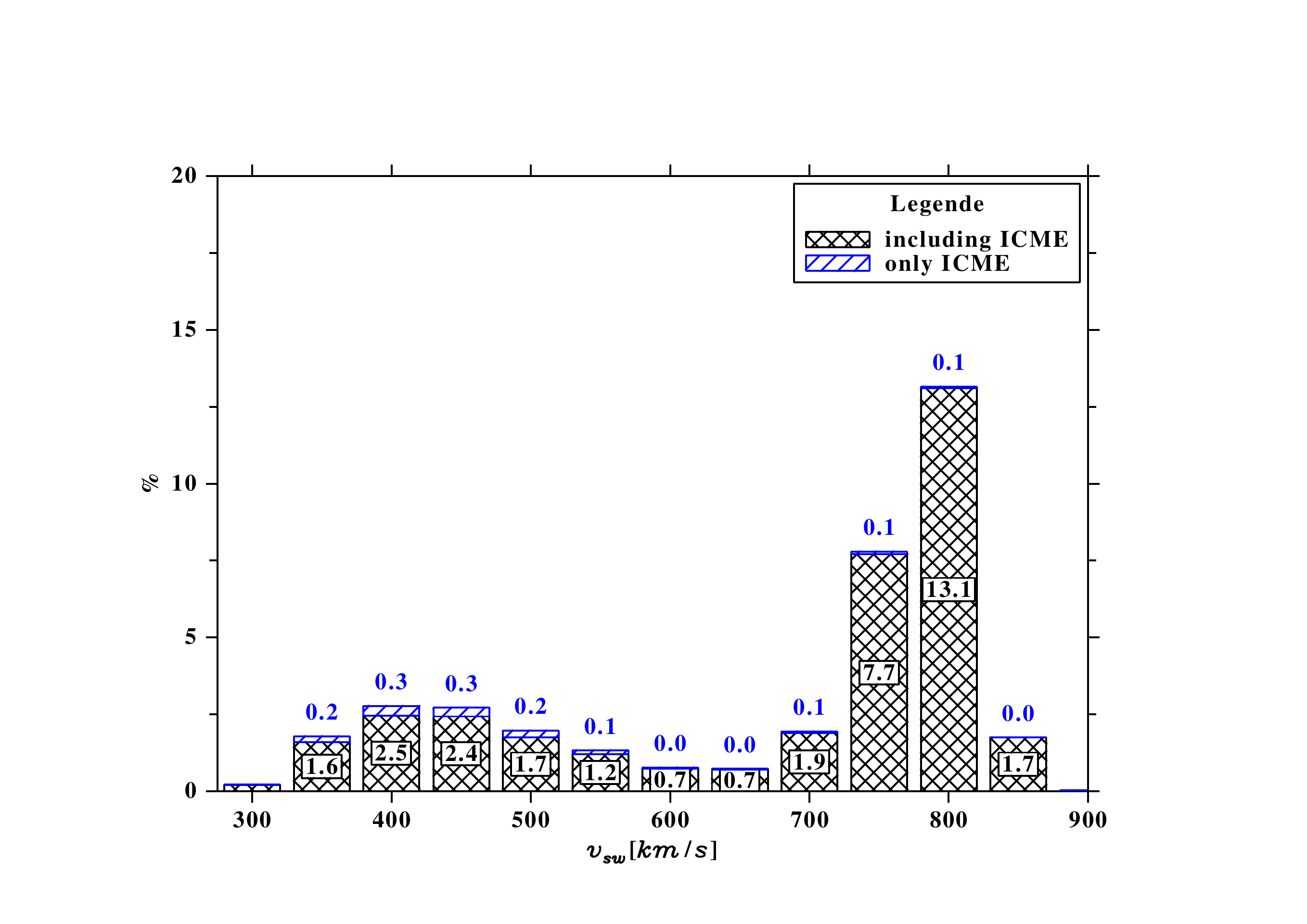}
  \caption{Histograms of the number of events depending on the solar wind
speed during the Ulysses mission: from low latitudes $\lt \ang{30}$ (upper panel) and from high latitudes $\gt \ang{40}$ (lower panel).  \label{fig:0b}}
\end{figure}

\section{Time and speed variations}\label{sec:time_speed_var}

We remind the reader of the time and latitude dependence of the solar wind speed along the Ulysses trajectory
\citep{McComas-etal-2008}, and reprint in Fig.~\ref{fig:0} the angular distribution from \citet{McComas-etal-2008}.
It can be seen that during the first and third latitude scan, there are low solar wind speeds below $\approx \ang{35}$
latitude and high speeds above $\approx \ang{35}$ latitude, and almost no intermediate speeds. This is different
during a more active Sun in the second scan, when the speeds scatter over all latitudes.

The difference in solar wind speeds depending on the latitude becomes also  evident in Fig.~\ref{fig:0b}, which shows
the histograms of the number of events (in \%) as a function of the solar wind speed.
We split into a part for low latitudes $\lt \ang{30}$ (upper panel) and a part for high latitudes $\gt \ang{40}$ (lower panel).
This time the events (and the corresponding values) without CMEs are given in black, while those during CMEs in blue.
The gap at intermediate speeds, with about 2.5~\% of events, is obvious around $\approx 650$\,\si{km/s}.  
From these histograms one can also observe that the low speeds cluster around 450\,km/s at low latitudes near the ecliptic (upper panel), while the high speeds cluster around 750\,km/s at high latitudes towards the poles and coronal holes (lower panel). Thus, high speeds are mainly observed during solar minimum \citep[see also][]{McGregor-etal-2011}. 

\begin{figure*}
    \centering
    \setkeys{Gin}{width=\linewidth}
    \begin{tabular}{cc}
    \begin{subfigure}[t]{\columnwidth}
      \includegraphics[height=0.3\textheight,valign=c]{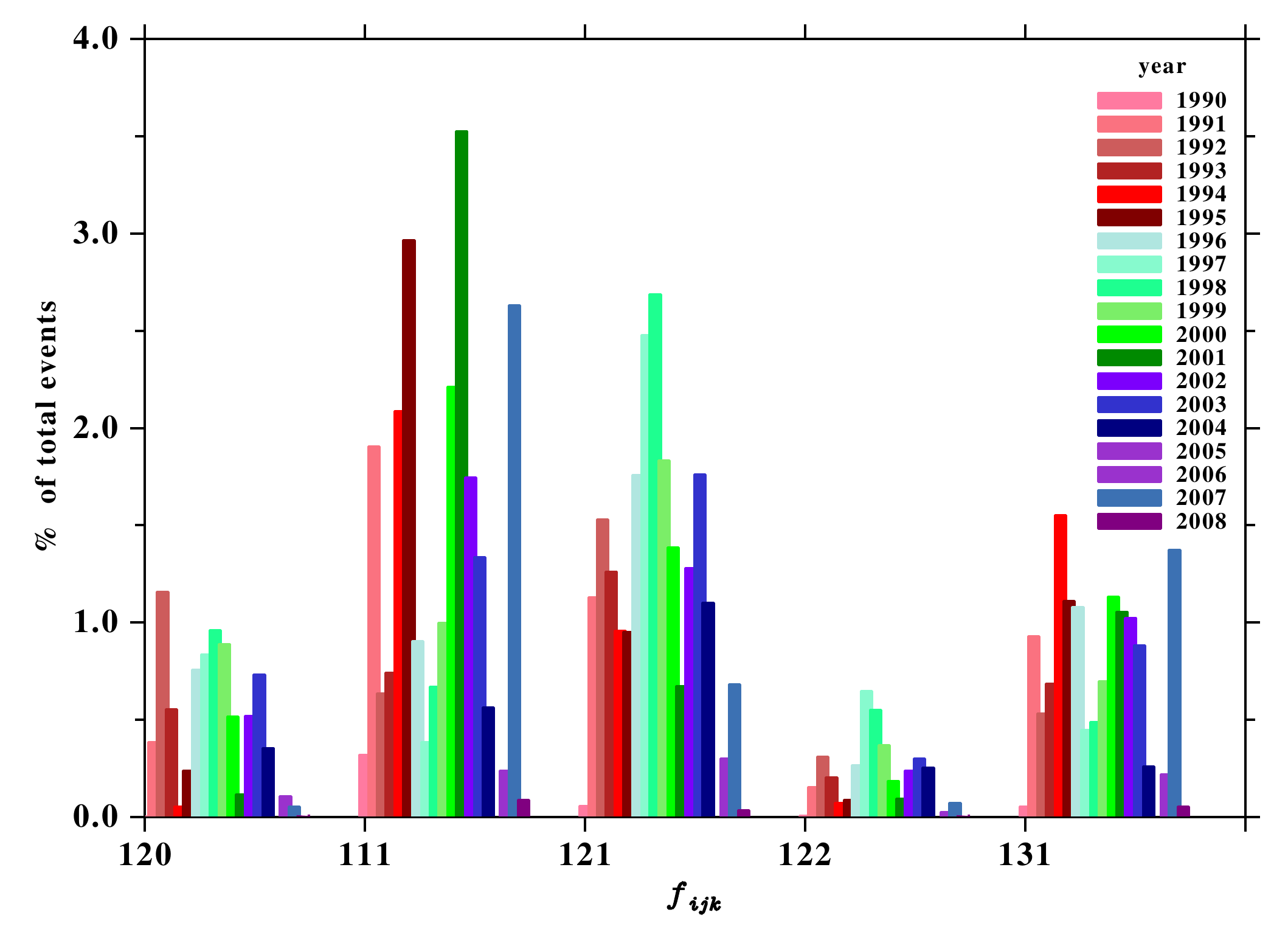}
    \end{subfigure}
    &
      \begin{subfigure}[t!]{0.95\columnwidth}
        \tiny
        \begin{tabular}{l|dddddd}
          year 
          & \multicolumn{1}{c}{$f_{120}$} 
          & \multicolumn{1}{c}{$f_{111}$} 
          & \multicolumn{1}{c}{$f_{121}$}
          & \multicolumn{1}{c}{$f_{122}$} 
          & \multicolumn{1}{c}{$f_{131}$}
          & \multicolumn{1}{c}{Sum} \\
          \hline
          1990 &  0.00 &  0.32 &  0.06 &  0.01 &  0.05 &  0.39 \\
          1991 &  0.38 &  1.90 &  1.13 &  0.15 &  0.93 &  3.72 \\
          1992 &  1.16 &  0.63 &  1.53 &  0.31 &  0.53 &  3.94 \\
          1993 &  0.55 &  0.74 &  1.26 &  0.20 &  0.68 &  2.96 \\
          1994 &  0.05 &  2.08 &  0.96 &  0.07 &  1.55 &  3.24 \\
          1995 &  0.24 &  2.97 &  0.95 &  0.09 &  1.11 &  4.33 \\
          1996 &  0.76 &  0.90 &  1.76 &  0.26 &  1.08 &  3.95 \\
          \hline
          $\sum^{1996}_{1990}$ & 3.14 & 9.54 &  7.65 & 1.87 & 5.94 & 22.44\\ 
          \hline
          1997 &  0.83 &  0.38 &  2.48 &  0.64 &  0.45 &  4.98 \\
          1998 &  0.96 &  0.67 &  2.68 &  0.55 &  0.49 &  5.41 \\
          1999 &  0.89 &  1.00 &  1.83 &  0.37 &  0.70 &  4.46 \\
          2000 &  0.52 &  2.21 &  1.38 &  0.19 &  1.13 &  4.48 \\
          \hline
          $\sum^{2000}_{1997}$ & 3.20 & 2.64 & 8.37  & 1.75  & 2.77  & 19.33\\ 
          \hline
          2001 &  0.12 &  3.52 &  0.67 &  0.10 &  1.06 &  4.50 \\
          2002 &  0.52 &  1.75 &  1.28 &  0.24 &  1.02 &  4.01 \\
          \hline
          $\sum^{2002}_{2001}$ & 0.64 & 5.27  & 1.95  & 0.34  & 2.08 & 8.51\\ 
          \hline
          2003 &  0.73 &  1.34 &  1.76 &  0.30 &  0.88 &  4.43 \\
          2004 &  0.35 &  0.56 &  1.10 &  0.25 &  0.26 &  2.52 \\
          2005 &  0.00 &  0.00 &  0.00 &  0.00 &  0.00 &  0.00 \\
          2006 &  0.11 &  0.24 &  0.30 &  0.03 &  0.22 &  0.70 \\
          2007 &  0.05 &  2.63 &  0.68 &  0.07 &  1.37 &  3.51 \\
          2008 &  0.00 &  0.09 &  0.03 &  0.00 &  0.05 &  0.13 \\
          \hline
          $\sum^{1008}_{2003}$ & 1.24 & 4.86  & 3.87  & 0.65  & 2.78  & 11.29\\ 
        \end{tabular}
        \label{tab:2}
\end{subfigure}
    \end{tabular}
    \caption{The histogram (left panel) and table of relative numbers (left panel) splitted into a yearly basis (right panel).
Left panel: Histogram for the BFs of individual years. Inside a distribution   function $f_{ijk}$ each bar represents the events per year, from 1990 to 2008. The data for 2005 are missing. The data for 1990 at the beginning of the mission are very sparse, which is also the case toward the end of the mission (after 2005). Right panel:
    The relative number of events 
by the BFs of the five most representative distribution functions. For each 
intermediary row, "$\sum^{yy}_{yy}$" gives the sum of the years above, and the 
last column "Sum" gives the sums of each row. The sum of the last 
column is 61.57\%. Thus, with these five distribution functions we cover 
approximately 62\% of all events (70.7\% events without CME).\label{fig:3}}
\end{figure*}
In Fig.~\ref{fig:3} the histogram of Fig.~\ref{fig:2} is divided in individual years. This helps us to find the combinations of distribution functions relevant for each orbit of the Ulysses missions, and implicitly for
different solar activities. These combinations are also indicated in Fig.~\ref{fig:0}, in order of their
relevance as follows: [$f_{111}$, $f_{121}$, $f_{131}$, $f_{120}$] for the first and third orbits, and [$f_{121}$, $f_{111}$, $f_{131}$, $f_{120}$] for the second orbit.
On the other hand, the $f_{111}$ and $f_{131}$ distributions have a dip around the years 1996 to 2000, which is the
ascending phase of solar cycle~23 (see Fig.~\ref{fig:0}), while the $f_{121}$
and $f_{122}$ distributions show a maximum. Unfortunately, that is the only ascending
phase which is covered by the Ulysses mission. Therefore, we can only
guess that during the ascending phases of a solar cycle, the $f_{121}$
($f_{122}$)
distributions are more appropriate. 
In two declining phases (that of solar cycle 22 and 23) the distribution functions  $f_{111}$ and $f_{131}$ give the most
relevant best fits, although $f_{121}$ is also well represented in this case. Thus, in a declining phase of a solar cycle, the
electron distributions are better reproduced by three Maxwellians, meaning that they are, individually, closer to thermal equilibrium.
Contrarily, in the rising phase, when the solar activity increases, the halo distribution is not well fitted by
Maxwellians, meaning that the particles are no longer in thermal equilibrium. The above holds also true for the period around 2002,
where the $f_{111}$ and $f_{131}$ distributions have a minimum, and later at 2007 show a maximum (and vice versa for
the $f_{121}$ and $f_{122}$ distributions). However, because this is at the end of the misson and no further data are available, we
cannot safely conclude that this time dependence is verified by observations. The reason is that the time series only covers parts of a solar Hale cycle, but needed were at least a few such cycles.
From a theoretical point of
view one may explain the time dependence by the changing magnetic field
in the rising phase and the beginning of the solar activity. Thus, the
non-equilibrium distributions $f_{121}$
($f_{122}$)
can be caused by an enhanced flare activity. This proposed context
needs a more detailed analysis including all spacecraft and solar
data, which is far beyond the topic of this work.

\section{Case study: temperature anisotropy} \label{sec:parameters}

\begin{figure*}[t!]
 \sidecaption
   \includegraphics[width=12cm]{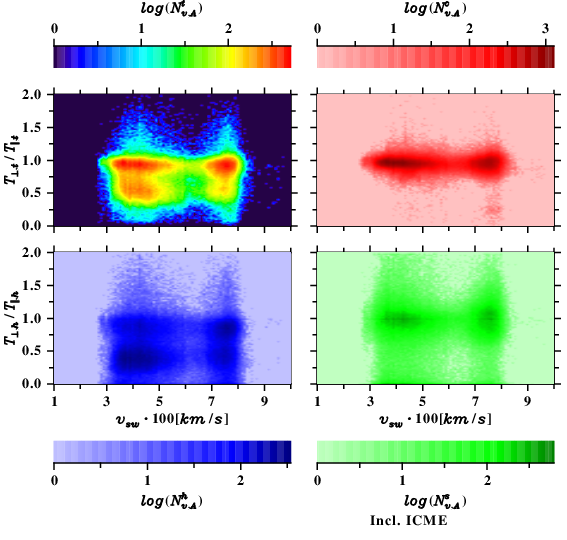}
   \caption{Correlating the solar wind speed (x-axis) given in units of 100\,km/s and
   the temperature anisotropy (y-axis) for all events. The number of events $N_{vA}^{i}$ is given in a logarithmic
   scale. Here we show the BFs of all fits, while in the following we also show the fits for $f_{111}, f_{121}$, and $f_{131}$ distributions. The upper-left panel shows the total anisotropy, the upper-right panel shows that of the core, while the lower-left and lower-right panels that of the halo and strahl, respectively.\label{fig:6}}
 \end{figure*}

\begin{figure*}[t!]
 \sidecaption
  \includegraphics[width=12.5cm]{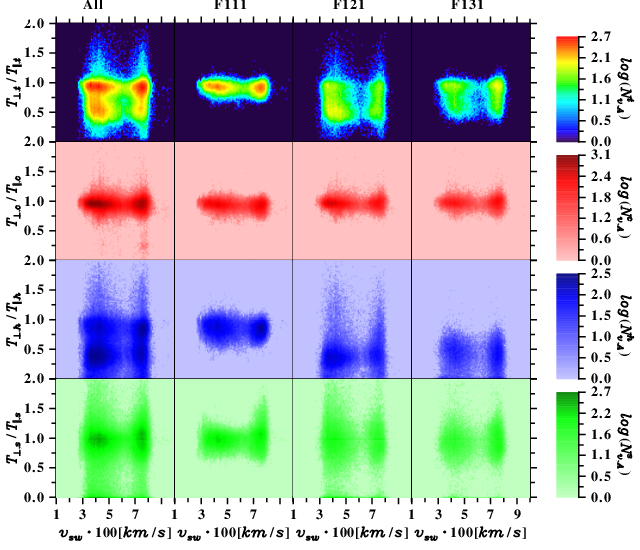}
   \caption{Correlating the solar wind speed and the temperature anisotropy without CMEs. The results shown in the columns from left to right are for $f_{all}$, $f_{111}$, $f_{121}$ and $f_{131}$, respectively. Rows from top to bottom display the results for the total, core, halo and strahl distributions, respectively. See text for more details.\label{fig:9}}
 \end{figure*}

\begin{figure*}[t!]
\sidecaption  
\includegraphics[width=12cm]{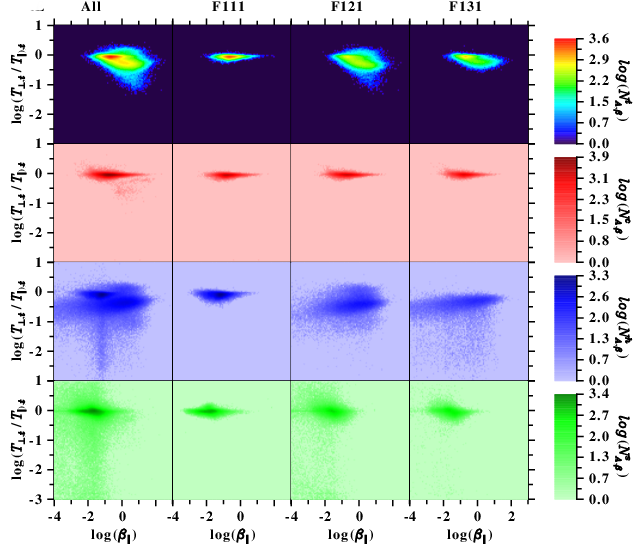}
   \caption{Correlating the parallel plasma beta and the temperature anisotropy. 
    Panels are organized as in Fig.~\ref{fig:9}. See text for more details\label{fig:b12}}
\end{figure*}

The four temperature anisotropies  we define as
\begin{align}\label{eq:a}
  A_{i} = \frac{T_{\se,i}}{T_{\pa,i}} = \frac{P_{\se,1}}{P_{\pa,i}}\,,\qquad i
  \in\{t,c,h,s\}
\end{align}
and additionally, we define the parallel plasma beta as
\begin{align}
  \beta_{\parallel,i} = \frac{8\pi P_{\parallel,i}}{B^{2}}, \qquad i
  \in\{t, c,h,s\}\,.
\end{align}
with magnetic field magnitude $B$.
The $\beta_\perp$ can be defined, but is not discussed here. See appendix~\ref{app:a} for further information including the representation method in Fig.~\ref{fig:6} to Fig.~\ref{fig:16}.
Furthermore, we only discuss the events without CMEs, and the distributions in the CMEs will be left for future work.
We display the total number of events on a rainbow colour-coded scale, while the number of core data is displayed in a reddish,
of the halo in a blueish, and of the strahl in a greenish color scale. 

\subsection{Temperature anisotropy and solar wind speed.}\label{subsec:temp_aniso_speed}

Figures~\ref{fig:6} and~\ref{fig:9} display the colour-coded binned number of events $N_{vA}^{i},\ i\in\{t,c,h,s\}$ (see
Sec.~\ref{sec:corr}), in a plot with  the solar wind speed versus the temperature anisotropy (Eq.~\ref{eq:a}). 
In Fig.~\ref{fig:6} we correlate temperature anisotropy and solar wind speeds, for the total distribution
function of all events, (left-upper panel), and only for the core (right-upper panel),
for the halo (left-lower panel), and for the strahl (right-lower panel). In the upper-left panel, four maxima in the number
of all events $N_{vA}^{t}$ can be identified: the first one between $v_{sw} = 300-500$\,km/s and an anisotropy of
$A\approx 1$, the second one between $v_{sw}\approx 700$\,km/s  and $v_{sw}\approx 800$\,km/s  and $A\approx1$, and the third and fourth at about the same
solar wind speeds, but at lower temperature anisotropies $A\approx 0.5$. The core distribution contributes mainly to the first
and second maxima (right-upper panel), while the halo distribution determines the third and fourth maxima, and contributes also to
the first and second maximum (left-lower panel). The strahl distribution (right-lower panel) contributes primarily to the first
and second maxima. This implies that the temperature anisotropy is mainly caused by the halo and strahl components, while the
core distribution is well fitted with an isotropic ($A=1$ temperature) distribution function. The dip at solar wind speeds about $v_{sw}\approx 650$\,\si{km/s} 
is due to the fact that these speeds are rare (see Figs.~\ref{fig:0} and ~\ref{fig:0b}). 

Figure~\ref{fig:9} is structured in a similar way as Fig.~\ref{fig:6} and shows a comparison of $f_{all}$ with $f_{111},
f_{121}$, and $f_{131}$ (the columns from left to right) by plotting $N_{vA}^{i}$ of the corresponding total, core, halo, and
strahl distributions. It can be seen that the $f_{111}$ distributions have maxima at temperature anisotropies around $A = 1$ for all
four plots: total, core, halo, and strahl. The $f_{121}$ distribution has two maxima along $A \approx 1$ and $A\approx 0.5$, 
where the former is mainly the contribution from the core, while the latter are contributions from the halo and the
strahl. The core of the $f_{121}$ distribution has a similar behavior as that of $f_{111}$ distribution, that is, it
scatters mainly around $A=1$. A similar behavior shows the $f_{131}$ distribution, except that the halo does not scatter as much as the halo of the $f_{121}$
distribution. Also, the total anisotropy is smoother for the $f_{131}$ distribution. The total anisotropy for the $f_{111}$
distribution scatters only around $A=1$ (temperature isotropy), while the total of the $f_{121}$ distribution has two maxima
around $A=1$ and $A=0.5$, similar to the $f_{131}$ distributions, except that the second maximum around $A=0.5$ is not very
pronounced. The dips are again explained by the absence of speeds about $v_{sw}\approx 650$\,\si{km/s} (see above).

The $f_{121}$ distribution shows the strongest scattering in the halo and superhalo component. 
Nevertheless, because this is the BF, it indicates that there might be another distribution function involved not
covered by the fitted AMD, RKD or GAK distributions. The scattering in the fitted halo by the $f_{121}$ distribution might
also occur due to the general nature of the GAK. We do not make a closer inspection here, but state that the bulk part of the
data is in the maxima of the $f_{121}$ distribution, so that the GAK can be used for fitting.

To conclude this section, we point out that the core has in this analysis a
bi-Maxwellian distribution, which can be replaced by an isotropic Maxwell
distribution, while the halo and superhalo/strahl are best fitted with anisotropic distributions (here an RAK or GAK). In many cases we can
also replace the GAK by the RAK (see the discussion in Sec.~\ref{sec:models}
above).

\subsection{Temperature anisotropy and parallel plasma beta} \label{subsec:temp_aniso_beta}

Figure~\ref{fig:b12} displays the temperature anisotropy as a function of the parallel plasma beta. 
In the top panel of first column the total temperature anisotropy of all
events is shown. For all
distributions, the core events (second row) show small deviations from isotropy and distribute regularly, more or less parallel to the $x$-axis. 
The halo events (first column, third row) predominantly show an excess of parallel temperature $T_{\parallel, h} > T_{\perp,h}$, revealing a mushroom-like shape, with the leg at $\beta_{\parallel}\approx 0.1$, and the hat widely ranging from very low to high values of $\beta_{\parallel, h}$, e.g.,  $10^{-4} < \beta_{\parallel, h} < 50$. The strahl events (first column, fourth
row) are similar, but 
spread at slightly lower values of $\beta_{\parallel, s} <1$. The distributions of these data are very similar to those obtained by \cite{Stverak-etal-2008} with the ecliptic electron data.
 
The second, third and fourth columns of Fig.~\ref{fig:b12} display the events fitted by $f_{111}$, $f_{121}$,
and $f_{131}$, respectively. It can be seen that the $f_{111}$
distribution has a very weak scattering in all components and is
quite similar to the above mentioned plots by
\citet{Stverak-etal-2008}. This is also true for the core events of
all other distributions (second row). However, the halo events of the
$f_{121}$ and $f_{131}$ distributions show much stronger scattering (third
row). Although more restrained, the strahl component (fourth row) show the same variation.  
For all distributions, the leg is thus a feature mainly resulting from the other distributions not shown here, e.g., $f_{120}$ and $f_{122}$ distributions.

\subsection{Temperature anisotropy and  $\kappa, \eta_{\pa}, \eta_{\se}$, and $\zeta$ parameters}\label{subsec:temp_aniso_parameters}

 \begin{figure*}  
 \sidecaption
   \includegraphics[width=12cm]{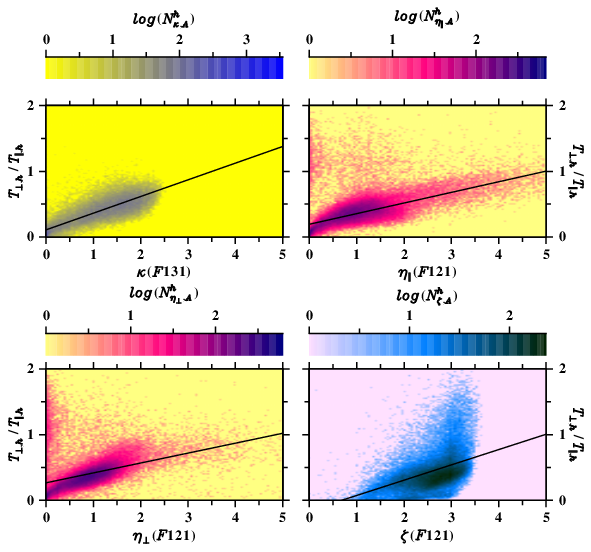}
   \caption{The correlation between the temperature anisotropy
   and parameters of the halo components of the distributions are displayed.
   The temperature anisotropy is compared with the $\kappa$ parameter of the
   $f_{131}$ distribution (left upper panel), and with the $\eta_{\pa}$, $\eta_{\se}$ 
   and $\zeta$ parameters of the $f_{121}$ distribution
     (right upper panel, left lower panel and right lower panel,
     respectively). For more explanation see text.
     \label{fig:16}}
 \end{figure*}

In Fig.~\ref{fig:16} we show the correlation between parameters of the halo components of
 the $f_{131}$ and $f_{121}$
 distributions with the temperature anisotropy. The color coding is now given at the top of each
 panel. In the upper left panel the scattering of the $\kappa$ values
 of the $f_{131}$
 distribution is shown. We can see that $\kappa$
 has only values between 0 to about 2.5.
 It is also evident that the lower the $\kappa$
 values are, the lower is the halo temperature anisotropy, that is, the
 perpendicular temperature is much higher than the parallel one. 
 The reason for $\kappa$
 values only below 2.5 can be due to the fact that for higher $\kappa$ values 
 the distribution is comparable to a Maxwellian, and thus the
 distribution $f_{121}$ approaches $f_{111}$. However, 
 this statement requires more research.

  For the $f_{121}$
  distribution, the scattering of $\eta_{\pa}$, $\eta_{\se}$
  and $\zeta$ parameters is similar to the scattering of $\kappa$ values in the $f_{121}$
  distribution, except that $\eta_{\pa}$
  has higher values (up to 5, see right upper panel), while
  $\eta_{\se}$ values have a similar range as $\kappa$
  (left lower panel). The $\zeta$ parameter (right lower panel) is similar to the $\eta_{\se}$ parameter. 
  The $f_{131}$ halo component shows also an increased scattering in the anisotropy around
  $\eta_{\pa}=0$, $\eta_{\se}=0$ and $\zeta=0$, which can be caused by
  the fitting procedure due to the small values of the corresponding parameters.

We computed a linear regression for the data shown in Fig.~\ref{fig:16} via
\begin{align}\label{eq:reg}
    A(x) = \frac{T_{\se h}}{T_{\pa h}} = a x + b\,,
\end{align}
where $x\in\{\kappa,\eta_{\pa},\eta_{\se}.\zeta\}$. The values for the fits are listed in Table~\ref{tab:fit}.
\begin{table}[t!]
    \caption{The fit values for the linear regression (Eq.~\ref{eq:reg})  as shown in Fig.~\ref{fig:16}.}
    \centering
    \begin{tabular}{l|r|r}
    $A(x)$ & a & b\\
    \hline
    $A(\kappa)$     & 0.25 & 0.11\\
    $A(\eta_{\pa})$ & 0.16 & 0.19  \\
    $A(\eta_{\se})$ & 0.15 & 0.27  \\
    $A(\zeta$)       & 0.23 & -0.16  
    \end{tabular}
    \label{tab:fit}
\end{table} 
The linear regression is quite good for $A(\kappa)$, but for $A(\eta_{\pa})$ and $A(\eta_{\se})$ the events form a curve asymptotically approaching $A=1$. The fit for 
$A(\zeta)$ runs through a cloud, which clusters around the linear regression and scatters toward higher anisotropies.
Nevertheless, the simplified regressions can help to study the temperature anisotropy in the halo with increasing values of $\kappa,\eta_{\pa},\eta_{\se}$ and $\zeta$.  

\section{Conclusions}\label{sec:conclusions}

In the present work we used a 2D fitting method, which accounts for different temperatures parallel and perpendicular
with respect to the background magnetic field, to fit electron velocity distributions obtained during the Ulysses mission. In doing so, 
we used a triple model to fit the core, halo and superhalo/strahl populations within the total distribution. As model functions we applied
an anisotropic Maxwellian distribution (AMD), a generalized anisotropic Kappa (GAK) and a regularized anisotropic Kappa (RAK) distribution. 
Our findings indicate
a time dependence of the electron distributions
on the solar cycle. Unfortunately, the data series is not
long enough to solidly confirm such a behavior. Nevertheless, the
results suggest that in a declining phase of a solar cycle, 
the individual electron distributions are best described by three Maxwellians, 
meaning that they are in thermal equilibrium. In the ascending phase, when the
solar activity increases, the halo distribution is not well fitted
by a Maxwellian, which means that the particles are out of thermal equilibrium. This
behaviour deserves more attention and should be be combined with other
spacecraft data. However, due to the unique solar polar orbit of Ulysses and the fact
that most of the other spacecraft are in the ecliptic (HELIOS) or
close to the Sun (Parker Space Probe, Solar Orbiter), one has in principle to subtract distance and latitude effects.    

While the core distribution can most likely be fitted with an isotropic Maxwell
distribution as the temperature anisotropy is close to 1 (see the reddish color panels in the figures in Sec.~\ref{sec:parameters}), the halo and superhalo/strahl (the blueish and greenish colo panels) are best fitted with
anisotropic distributions (here with an RAK or a GAK. In many cases we can
also replace the GAK by the simpler RAK, resulting in a slightly worse fit (see Fig.~\ref{fig:20}).

The parallel plasma beta correlation with the temperature anisotropy is similar for the three distributions $f_{111}, f_{121}$, and $f_{131}$, but shows a stipe for the halo distribution in the sample of all distributions.
We also showed the correlation between the temperature anisotropy and
the $\kappa$ parameter of the $f_{131}$ distribution as well as the
$\eta_{\pa}$, $\eta_{\se}$ and $\zeta$ parameters of the $f_{121}$
distribution. The temperature anisotropy shows almost a linear dependence on $\kappa$ with increasing
values of $\kappa$, and the temperature anisotropy becomes more isotropic. For the GAK parameters 
($\eta_{\pa},\eta_{\se},\zeta$) we found also a linear dependence of the temperature anisotropy.    

Finally, we want to direct the attention to Fig.~\ref{fig:0} which shows that during a solar cycle also the type of the distribution functions changes from mainly a $f_{111}$ type near solar minimum to a more general  $f_{121}$ type close to solar maximum and back to $f_{111}$ for the next minimum phase. 
These results demonstrate that the multi-distribution function fitting of velocity distributions has a significant potential to advance our understanding of the solar wind kinetics and, therefore, deserves further quantitative analyses.
 
The data sets were derived from sources in the public domain, available at \verb+http://ufa.esac.esa.int/ufa/#data/+.

\begin{acknowledgements} 
The authors acknowledge support from the Ruhr-University Bochum and the Katholieke Universiteit Leuven.  EH is grateful to the Space Weather Awareness Training Network (SWATNet) funded by the European Union’s Horizon 2020 research and innovation programme under the Marie Sk{\l}odowska-Curie grant agreement No 955620.
\end{acknowledgements}

\bibliographystyle{aa}
\bibliography{test}

\appendix
\section{The pressure and temperature moments}\label{app:a}

We repeat here shortly the definitions given in \citet{Paschmann-etal-1998} and \citet{Scherer-etal-2021}.
The partial parallel and perpendicular thermal pressures are given as follows:
\begin{align}
    P_{th} =& \frac{1}{3} \left(\sum\limits_{i} P_{\parallel,i} + 2\sum\limits_{i}
             P_{\perp,i} \right) \qquad i \in\{c,h,s\}\\\nonumber
  \Rightarrow &
  \begin{cases} 
    P_{\parallel,th}  \equiv \sum\limits_{i} P_{\parallel,i}\\
   P_{\perp,th}  \equiv \sum\limits_{i} P_{\perp,i}\,.
    \end{cases}
\end{align}
The total pressure $P_{tot}$
is the sum of the thermal pressure plus (twice) the ram pressures (see \cite{Scherer-etal-2021} for further explanation).
 
The respective temperatures are given by the ideal gas law as
\begin{subequations}
\begin{align}
    T_{\parallel,i} &\equiv \frac{P_{\parallel,i}}{k_{B} n_{i}}\,,\qquad \qquad 
    T_{\perp,i} \equiv \frac{P_{\perp,i}}{k_{B} n_{i}}\,,  \qquad i  \in\{c,h,s\}
\end{align}
with $k_B$ denoting Boltzmann's constant, and
\begin{align}
  T_{\parallel,t} = \sum\limits_{i}\frac{n_{i}T_{\parallel,i}}{n_{t}}\,,\qquad\qquad
  T_{\perp,t} = \sum\limits_{i}\frac{n_{i}T_{\perp,i}}{n_{t}}\,,
\end{align}
where number density $n_{t}$ is given by
\begin{align}
  n_{t} = \sum\limits_{i}n_{i}\,,\qquad i  \in\{c,h,s\}\,.
\end{align}

\end{subequations}
\subsection{Representation method} \label{sec:corr}

In the graphical representations we divide the $x$- and $y$-axis in 100 sub-intervals $\Delta x_{i}$ and $\Delta
y_{j}$, and compute the total number of events (data points) $N_{all}$  in  each interval $N_{i,j} / (\Delta x_{i} \Delta y_{j})$ via
\begin{align}
  N_{all} = \sum\limits_{i,j} N_{i,j} / (\Delta x_{i} \Delta y_{j})\,.
\end{align}
As stated above, we only discuss the macroscopic parameters for the most relevant distribution functions, according to their BFs, that is,
$f_{all},f_{111},f_{121}$, and $f_{131}$, where $f_{all}$ are all the BFs including all distributions $f_{ijk}$.
We always show first the total, core, halo and superhalo/strahl
moments for $f_{all}$,
and then we show how the moments for the single distributions
$f_{111},f_{121}$ and $f_{131}$
contribute to $f_{all}$.

\end{document}